\documentclass[12pt]{article}

\usepackage{amssymb}
\usepackage{color}
\usepackage{epsfig,amssymb,amsfonts,amsmath,graphicx,cite,xfrac}
\usepackage{authblk}
\usepackage{subcaption}
\definecolor{mygray}{gray}{0.5}

\usepackage{cite}
\usepackage[colorlinks=true,linkcolor=blue,citecolor=red]{hyperref}

\newcommand{\be}{\begin{equation}}
\newcommand{\ee}{\end{equation}}
\newcommand{\bea}{\begin{eqnarray}}
\newcommand{\eea}{\end{eqnarray}}

\title{Modeling quantum information dynamics achieved with time-dependent driven fields in the context of universal quantum processing}

\author[${}$]{Francisco Delgado}
\author[${}$]{Suset Rodr\'iguez}

\affil[${}$]{\footnotesize  Tecnol\'ogico de Monterrey, Escuela de Ingenier\'ia y Ciencias, Carretera Lago de Guadalupe Km. 3.5, Atizap\'an de Zaragoza, Estado de M\'exico, C.P. 52926, M\'exico}

\date{}
\begin{document}

\maketitle

\begin{abstract}
Quantum information is a useful resource to set up information processing. Despite physical components are normally two-level systems, their combination with entangling interactions becomes in a complex dynamics. Studied for piecewise field pulses, this work analyzes the modeling for quantum information operations with fields affordable technologically towards a universal quantum computation model.
\end{abstract}


\section{Introduction}
Quantum information takes advantage of quantum physical systems and their properties as superposition and entanglement, where two-level systems are combined and scaled becoming in complex behaviors. Instead of the original states of isolated systems, the use of more advisable non-local basis to set the grammar lets exceed the complexity (polarization or spin). We focus the analysis on magnetic systems for two qubits following the Heisenberg-Ising model with magnetic fields in a fixed direction $h=1, 2, 3$:

\begin{eqnarray} \label{hamiltonian}
H_h &=& \sum_{k=1}^3 J_k {\sigma_1}_k {\sigma_2}_k -{B_1}_h {\sigma_1}_h-{B_2}_h {\sigma_2}_h 
\end{eqnarray}

\noindent here, Bell states recover the binary dynamics \cite{delgadoalgebraic}. The $SU(4)$ dynamics is split in two subspaces with $SU(2)$ dynamics ($U(1) \rtimes SU(2)^2$ as $U_{h}(t)={s_h}_{1} \oplus {s_h}_{2}$). In terms of Hilbert space: $\mathcal{H}_h^{\otimes 2} = \mathcal{H}_{h,1} \oplus \mathcal{H}_{h,2}$, states are split in the subspaces $\left| \psi(t) \right> = \alpha_1 \left| \psi_1(t) \right> + \alpha_2  \left| \psi_1(t) \right>$, $|\alpha_1|^2 + |\alpha_2|^2 = 1$ and $\left| \psi_k(t) \right> = {s_h}_{k} \left| \psi_k(0) \right>$. Then, $H_h = H_{1,h} \oplus H_{2,h}$ \cite{delgadoalgebraic} with:

\begin{align}\label{hblocks}
H_{k,h} \equiv \tilde{H}^0_{k,h} + \tilde{H}_{k,h} \rightarrow
\tilde{H}^0_{k,h} = -s_0 J_h \sigma_0^{(h,k)}, \quad \tilde{H}_{k,h} = s_1 {J_{\{h\}}}_{s_0} \sigma_3^{(h,k)} + s_2 {B_h}_{-s_0} \sigma_{q}^{(h,k)} 
\end{align}

\noindent $\sigma_i, i=0,...,3$ are the Pauli matrices for a space generated by paired Bell states. $\sigma_{q}^{(h,k)} = - (q-2) \sigma_1^{(h,k)} + (q-1) \sigma_2^{(h,k)}$, with $s_0=(-1)^{h+k+1}, s_1=s_0^{p}, s_2=(-1)^{p} s_0^{p+q}$ and $p=1+\frac{1}{2}(h-1)(h-2), q=2- h {\rm mod} 2$ depending on $h, k$, together with the arrangement of pairs of Bell states generating each subspace \cite{delgadoalgebraic} (not relevant for this development). $J_h, {J_{\{h\}}}_{s_0}, {B_h}_{-s_0}$ are obtained from strengths and fields. The block structure of  $\tilde{H}_{k,h}$ for $k=1, 2$ is inherited to $U_h(t)={s_h}_1 \oplus {s_h}_2$ through the time ordered integral \cite{gross} fulfilling the Sch\"odinger equation. As $\tilde{H}^0_{k,h}$ commutes with $\tilde{H}_{k,h}$, by defining ${s_h}_{k} \equiv \exp({-\frac{i}{\hbar}}\tilde{H}^0_{k,h}){s_h}^0_{k}$ (revealing the $U(2)=U(1) \times SU(2)$ structure), we get ${s_h}^0_{k}$ fulfilling the Sch\"odinger equation with $\tilde{H}^0_{k,h}$. Then, we will work with $\tilde{H}_{k,h}$ and ${s_h}^0_{k}$. Analytical solutions for time-independent or stepwise fields exist \cite{delgadocontrol, delgadogates} but they are few feasible because resonant effects. 

This work sets a procedure a control procedure with fields as those in resonant cavities, ion traps and laser beams \cite{seri1, brit1, bohn1, neto1}. In the second section, we benchmark linear and quadratic models to solve numerically the time-dependent problem. Third section  prescriptions to reduce the evolution into customary processing operations. At the end, we conclude how these results contribute for feasible models of experimental control. 

\section{Evolution in the time-dependent regime}

Baker-Campbell-Hausdorff formula rarely provides closed analytical solutions for time-dependent problems. Alternatively, numerical approaches are necessary. Here, we combine the $SU(2)$ reduction with linear and quadratic approaches to solve the time-dependent problem getting the generic block ${s_h}_j$ for composite quantum systems. A comparative benchmark is presented at the end. By defining the differential evolution operator $\left| \psi_k(t_0 + \delta t) \right> = {s_h}^0_{k}(t_0 + \delta t,t_0) \left| \psi_k(t_0) \right>$ and $J_h=\hbar{\mathcal J}_0, -s_1 {J_{\{h\}}}_{s_0}=\hbar{\mathcal J}_{s_0}, -s_2 {B_h}_{-s_0}=\hbar{\mathcal B}_{-s_0}$:

\begin{eqnarray} \label{dsh}
{s_h}^0_{k}(t_0 + \delta t,t_0) &\approx& {\bf 1}_k - \frac{i}{\hbar} \tilde{H}_{k,h} (t_0) \delta t - \frac{i}{2\hbar} \left( \frac{\partial \tilde{H}_{k,h}(t_0)}{\partial t} + \frac{\tilde{H}^2_{k,h}(t_0)}{i\hbar} \right) \delta t^2 + ... \\ 
&\approx& \sigma_0^{(h,k)} + i L^{(h,k)} \delta t - \frac{1}{2} Q^{(h,k)} \delta t^2 + ... 
\end{eqnarray}

\noindent with: $L^{(h,k)}={\mathcal J}_{s_0}  \sigma_3^{(h,k)} + {\mathcal B}_{-s_0}(t_0) \sigma_{q}^{(h,k)}$ and $Q^{(h,k)}=({\mathcal J}_{s_0}^2 + {\mathcal B}^2_{-s_0}(t_0)) \sigma_0^{(h,k)} - i {\mathcal B}'_{-s_0}(t_0) \sigma_{q}^{(h,k)}$. By splitting $[0, t]$ in $n$ intervals $[0, \delta t] \cup ... \cup [(n-1) \delta t, n \delta t = t]$ and using (\ref{hblocks}):

\begin{eqnarray} \label{dsh2}
{s_h}_{k} &=& {s_h}_{k}(t,0) \approx {\prod_{i=1,...,n}^\leftarrow} {s_h}_{k}( i \delta t, (i-1) \delta t) = e^{i s_0 {\mathcal J}_0 t} {\prod_{i=1,...,n}^\leftarrow} {s^0_h}_{k}( i \delta t, (i-1) \delta t) 
\end{eqnarray}

\noindent where $\leftarrow$ means factors stack on the left and $\delta t \approx 0$ is assumed. As in the independent-time case, ${\mathcal J}_0$ \cite{delgadocontrol} is only responsible from the weak $U(1)$ link between the two $SU(2)$ blocks through $e^{i s_0 {\mathcal J}_0 t}$. (\ref{dsh}) is a second order approximation in $\delta t$ for ${s_h}_k$. We have reduced the dependence ${s_h}_{k}$ ($k=1,2$) on the parameters ${\mathcal J}_0, {\mathcal J}_\pm, {\mathcal B}_{\pm}(t)$. 

\section{Construction of universal operations}

We fix ${B_h}_\pm(t)$ in a model applicable for resonant cavities, laser beams, ion traps or superconducting circuits \cite{seri1, brit1, bohn1, neto1}. Despite possible modes ($m=1, 2, ...$), we will take one single mode $m$. Absorbing physical quantities: $t'=\frac{mc}{d}t$ ($d$ an effective wavelength or length in the system) and ${\mathcal A}'_\pm = {\mathcal A}_\pm \frac{d}{mc}, {\mathcal J}'_\pm = {\mathcal J}_\pm \frac{d}{mc}$, we assume (dropping the apostrophe): 

\begin{eqnarray} \label{field0}
{B_h}_\pm(t) &=& {\mathcal A}_\pm \sin(\pi t)
\end{eqnarray}

For the numerical approach of ${s_h}_{k}$, we develop a benchmark comparing linear and quadratic approximations in (\ref{dsh}, \ref{dsh2}) (Figure 1a). By running $5 \times 10^4$ random experiments distributed uniformly on ${\mathcal J}_\pm, {\mathcal A}_{\pm}$, we track both the average time of computer processing and the $p$ correct figures reached. The process was repeated for $n$ ranging from $10$ to $10^4$, showing at least one figure of improved performance for the quadratic approach in (\ref{dsh}), then reducing $n$ from thousands to hundreds. Our implementation (second order and $n=100$) reaches at least five figures of precision. Introducing (\ref{field0}) in (\ref{dsh}), we are interested in the reduction of ${s_h}_{k}$ into:

\begin{figure}[ht] \label{fig1}
\begin{center}
\setlength{\abovecaptionskip}{-12pt}
\includegraphics[width=38.5pc]{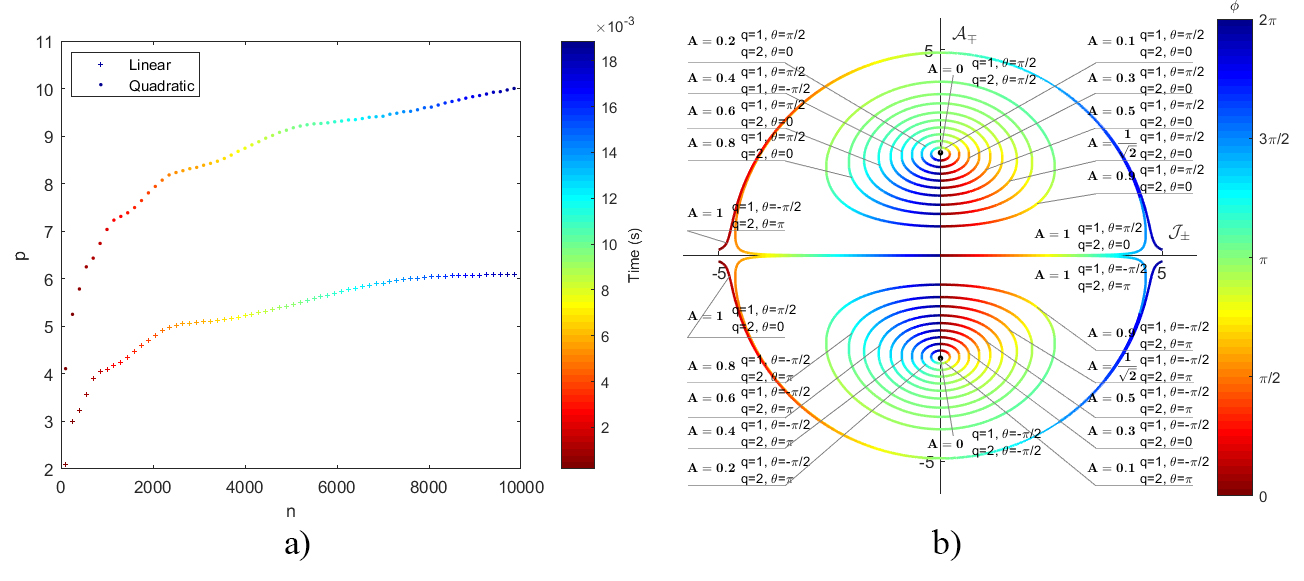}\hspace{1pc} 
\caption{a) Benchmarking for linear and quadratic approximation as function of $n$ (partition), $p$ (digits) and time computer processing (Intel Core i7 2.6GHz); and b) Solutions for ${\mathcal A}_{\mp}, {\mathcal J}_{\pm}$ with $\phi$ in the color chart and $\theta$ as an inset for a set of values for $A$.}
\end{center}
\end{figure}

\begin{eqnarray} \label{generic}
{s_h}_{k} = e^{i \varphi} {s_h}^0_{k} \equiv e^{i \varphi} { \left(
\begin{array}{cc}
A e^{i \phi} & B e^{i \theta}   \\
-B e^{-i \theta} & A e^{-i \phi}   
\end{array}
\right) }
\end{eqnarray}

\noindent with $A^2 + B^2 =1$. It reduces the dynamics in several traditional quantum processing operations as $C^aNOT_b$, $NOT$, Hadamard, etc. under the $SU(2)$ reduction scheme \cite{delgadogates}. It is affordable imposing restrictions on $A$, and some concrete prescriptions for $\phi, \theta$ and $\varphi$ during $t \in [0,1]$.  Figure 1b shows a set of solutions for $A=0, 0.1, ...,1$ in the region $[-5, 5] \times [-5, 5]$ of ${\mathcal A}_{\pm}, {\mathcal J}_{\mp}$ plane for $q=1, 2$. Insets specify the values of $A, q$ and $\theta$ (unique for each curve) and $\phi$ is reported in the color chart.

\section{Conclusions}
Solutions obtained for the generic processing operations could be combined under the scheme proposed in \cite{delgadogates} for affordable fields as (\ref{field0}). The numerical process developed shows to be efficient to obtain the prescriptions. Then, multi-mode implementations are possible for a common period or otherwise the semi-pulse presented here could be obtained as a superposition of infinite modes as a Fourier series in the set-ups depicted before, letting establish a sequence of semi-pulses corresponding to different gates one followed by another.


\section*{Acknowledgment}

The support of CONACyT and Tecnol\'ogico de Monterrey is acknowledged.

\end{document}